\documentstyle[12pt, epsf]{article}
\def\ga{\mathrel{\raise.3ex\hbox{$>$\kern-.75em\lower1ex\hbox{$\sim$}}}}
\def\la{\mathrel{\raise.3ex\hbox{$<$\kern-.75em\lower1ex\hbox{$\sim$}}}}
\def\gev{{\rm \, Ge\kern-0.125em V}}
\def\tev{{\rm \, Te\kern-0.125em V}}
\def\beq{\begin{equation}}
\def\eeq{\end{equation}}
\def\ss{\scriptscriptstyle}
\def\mb{m_{\widetilde B}}
\def\msf{m_{\tilde f}}

\def\mfb{\overline{m}_{\ss f}}
\def\mf{m_{\ss{f}}}

\def\gf{\gamma_f}
\def\thm{\theta_\mu}
\def\tha{\theta_A}

\def\cp{C\!P}
\def\ohsq{\Omega_{\widetilde\chi}\, h^2}

\begin{document}
\begin{titlepage}
\pagestyle{empty}
\baselineskip=21pt
\rightline{UMN--TH--1423/96}
\rightline{TPI--MINN--96/2}
\rightline{February 1996}
\vskip1.25in
\begin{center}

{\large{\bf Electric Dipole Moment Constraints
on Phases in the Constrained MSSM}}
\end{center}
\begin{center}
\vskip 0.5in
{Toby Falk and Keith A.~Olive
}\\
\vskip 0.25in
{\it
{School of Physics and Astronomy,
University of Minnesota, Minneapolis, MN 55455, USA}\\}
\vskip 0.5in
{\bf Abstract}
\end{center}
\baselineskip=18pt \noindent
%%%%%%%%%%%%%%%%%%%%%%%%%%%%%%%%%%%%%%%%%%%%%%%%%%%%%%%%%%%%%%%%%%%%%
We consider constraints on $\cp$-violating phases in the
Constrained Minimal Supersymmetric Standard Model.  We find that
by combining cosmological limits on gaugino masses with experimental
bounds on the neutron and electron electric dipole moments, we can
constrain the phase of the Higgs mixing mass $\mu$ to be $|\thm|
< \pi/10$, independent of choices of the other mass parameters in
the model.  The other $\cp$-violating phase $\tha$ is essentially
unconstrained.
%%%%%%%%%%%%%%%%%%%%%%%%%%%%%%%%%%%%%%%%%%%%%%%%%%%%%%%%%%%%%%%%%%%%%
\end{titlepage}
%\newpage
\baselineskip=18pt
%%%%%%%%%%%%%%%%%%%%%%%%%%%%%%%%%%%%%%%%%%%%%%%%%%%%%%%%%%%%%%%%%%%%%
The difficulties associated with trying to constrain the vast
parameter space of the general low-energy Minimal Supersymmetric
Standard model are well known, and some simplifying assumptions are
usually be made in order to get an experimental handle on the set of
new parameters.  A common ans\"{a}tz is that of the Constrained Minimal
Supersymmetric Standard Model, based on supergravity and grand
unification (described in more detail below).  The CMSSM is
interesting because it is simple, predictive and naturally provides a
stable dark matter candidate \cite{susygut} (an LSP bino-type
neutralino) over most of its parameter space. In \cite{fkos}, we
showed that there was a strong correlation between the $\cp$-violating
phases in the MSSM, the cosmological LSP relic density and the neutron
and electron electric dipole moments, when all of the
parameters are set at the weak scale and hence are independent of any RGE
evolution. Here we wish to consider the effect of the $\cp$-violating
phases present (though normally ignored) in the CMSSM, specifically
the inducing of electric dipole moments for the neutron and the
electron. We will find that by combining cosmological constraints on
the mass of the LSP with experimental bounds on the neutron and
electron EDM's, we can bound one of the two new phases in the CMSSM to
lie within $|\thm|<\pi/10$.

The MSSM contains a plethora of new parameters, which make empirically
testing the model difficult.  In addition to the supersymmetric Higgs
mixing mass parameter $\mu$, there are supersymmetry-breaking gaugino
masses $M_i$, sfermion mass$^2$ parameters $m_{\tilde f_{\ss i}}^2$,
trilinear scalar parameters $A_i$, the Higgs scalar mixing mass$^2\;
B\mu$, and the ratio of the Higgs vevs, $\tan\beta$.  This large
parameter space can be simplified at the unification scale by taking a
common gaugino mass $M$, sfermion mass$^2$ parameter $m_0^2$ and
trilinear scalar parameter $A$, and this model is referred to as the
CMSSM.  Since in the CMSSM, the two Higgs mass$^2$ parameters are set
equal to the sfermion mass$^2$ parameters at $M_X$, the other Higgs
sector masses are determined by the requirement of correct electroweak
symmetry breaking, once $\tan\beta$ and $m_{\rm top}$ are fixed.  In
principle, $M, \mu, B\mu, $ and $A$ may be complex; however, not all
of these phases are physical \cite{dgh}. It is possible to rotate away
the phase of the gaugino masses. And, by making $B\mu$ real, we ensure
that the vacuum expectation values of the Higgs fields are real.  The
CMSSM, then, is specified by three masses ($m_0, M, $ and $A_0$), two
phases ($\thm (M_X)$ and $\tha (M_X)$), $\tan\beta$ and $m_{\rm top}$.

Once the masses and phases are given at $M_X$, they can be RGE evolved
to the electroweak scale.  In practice, we use the one-loop RGEs for
the masses and two-loop RGEs for the gauge and Yukawa couplings
\cite{ikkt}. The structure of the equations for the gauge couplings,
gaugino masses and the diagonal elements of the sfermion masses are
such that they are entirely real.  The evolutions of the $A_i$,
however, are more complicated, as the $A_i$ pick up both real and
imaginary contributions.  For example, the evolution of $A_t$ is given
by
\begin{eqnarray}
  {dA_t \over dt} = {1\over{8\pi^2}}\left(-{16\over3}\,g_3^2\, M_3
    -3g_2^2\, M_2-{13\over9}\, g_1^2\, M_1 + h_b^2 A_b + 6 h_t^2 A_t\right)
\end{eqnarray}
As one can see, $A_t$ receives real contributions $c_i M$ proportional
to the gaugino mass (whose coefficients $c_i$ are different for each
sfermion in a generation) and (in principle complex) contributions
$d_i h_f^2 A_f$ from the heavy generation (whose coefficients $d_i$
differ for the first two and the third generations); the phases (and
magnitudes) of the $A_i$ must therefore be run separately.  At one
loop, the evolution equation for $\mu$ is given by
\begin{eqnarray}
  {d\mu \over dt} = {\mu\over{16\pi^2}}\left(-3g_2^2-g_1^2+h_\tau^2+
    3h_b^2+3h_t^2\right)
\end{eqnarray}
and the phase of $\mu$ does not run.

The aim of this paper is to combine cosmological bounds on the relic
density of neutralinos with experimental bounds on the neutron and
electron electric dipole moments in order to place limits on the new
$\cp$-violating phases $\thm$ and $\tha$.  The relic cosmological
density of neutralinos and the electric dipole moments are both
strongly dependent on the sfermion masses.  We take the general form
of the sfermion mass$^2$ matrix to be \cite{er}
\begin{equation}
\pmatrix{ M_L^2 + m_f^2 + \cos 2\beta (T_{3f} - Q_f\sin ^2 \theta_W) M_Z^2 &
m_f\,\overline{m}_{\ss f} e^{i \gamma_f}
\cr
\noalign{\medskip} m_f\,\overline{m}_{\ss f} e^{-i \gamma_f} & M_R^2 + m_f^2 +
\cos 2\beta Q_f\sin ^2
\theta_W M_Z^2
\cr }~
\end{equation}
where $M_{L(R)}$ are the soft supersymmetry breaking sfermion mass
which we have assumed are generation independent and generation
diagonal and hence real.  Due to our choice of phases, there is a
non-trivial phase associated with the off-diagonal entries, which we
denote by $\mf(\overline{m}_{\ss f} e^{i \gamma_f})$, of the sfermion
mass$^2$ matrix, and
\begin{equation}
  \label{mfbar}
  \overline{m}_{\ss f} e^{i \gamma_f} = R_f \mu + A_f^* = R_f
  |\mu|e^{i \theta_\mu} + |A_f|e^{-i \theta_{A_{\ss f}}},
\end{equation}
where $m_f$ is the mass of the fermion $f$ and $R_f =
\cot\beta\:(\tan\beta)$ for weak isospin +1/2 (-1/2) fermions.  We
also define the sfermion mixing angle $\theta_f$ by the unitary matrix
$U$ which diagonalizes the sfermion mass$^2$ matrix,
\begin{equation}
U = \pmatrix{ \cos\theta_{\!f} & \sin\theta_{\!f}\, e^{i\gamma_f} \cr
  \noalign{\medskip}
              -\sin\theta_{\!f} \,e^{-i\gamma_f} & \cos\theta_{\!f}\cr }.
\end{equation}

Previously \cite{fkos}, it has been shown that the presence of new
$\cp$-violating phases may have a significant effect on the relic
density of bino-type neutralinos.  The dominant channel for bino
annihilation is into fermion anti-fermion pairs.  However, this
process exhibits p-wave suppression, so that the zero temperature
piece of the thermally averaged annihilation cross-section (which is
relevant for the annihilation of cold binos) is suppressed by a factor
of the final state fermion mass$^2$.  This significantly reduces the
annihilation rate and increases the neutralino relic density.  Mixing
between left and right sfermions lifts this suppression to some extent
by allowing an s-wave contribution to the annihilation cross-section
which is proportional to the bino mass$^2$, but the presence of
complex phases in the off-diagonal components of the sfermion mass
matrices dramatically enhances this effect. Explicitly, we compute the
relic density by using the method described in ref.~\cite{swo} and
expand $\langle \sigma v_{ rel } \rangle$ in a Taylor expansion in
powers of $T/ m_{ \widetilde B } $
\begin{equation}
 \langle \sigma v_{ rel } \rangle = a + b\:({T/\mb}) +
O\left(\left({T / \mb}\right)^2\right)
\end{equation}
The coefficients $a$ and $b$ are given by
\begin{equation}
a = \sum_f v_f \tilde a_f
\end{equation}
\begin{equation}
b = \sum_f v_f \left[ \tilde b_f + \left( -3 + {3 m_f^2 \over 4 v_f^2
    m_{\widetilde B}^2 } \right) \tilde a_f \right]
\end{equation}
where $ \tilde a_f $ and $ \tilde b_f $ are computed from the expansion of the
matrix element squared in powers of $p$, the incoming bino momentum, and
$v_f = (1 - m_f^2/m_{\widetilde B}^2)^{1/2}$
is a factor from the phase space integrals.  For $M_L^2\approx M_R^2$ and
$m_f \ll \mb$ \cite{fkos},
\begin{equation}
  \label{af}
  {\tilde a}_f \;\approx \; {{(g')}^4\over 32\pi}\:
{\mb^2\over (\msf^2+\mb^2-m_f^2)^2}\: Y_L^2\: Y_R^2\:
\sin^2(2\theta_f)\sin^2\gamma_f
\;\;+ \;\;O(\mf\mb)
\end{equation}
We see that if sfermion mixing is significant ($\theta_f$ is large) and the
phase of the mixing term is large, the p-wave suppression is removed.

The effect of $\cp$-violating phases in a simple model where all the
sfermion scalar mass parameters were (for convenience) taken equal at
the electroweak scale, along with the trilinear $A$ parameters, was
studied in \cite{fkos}.  It was found that the cosmological upper
bound on the bino mass in this model was increased from $250\gev$
\cite{osi3,gkt} to $650\gev$ by the presence of the new phases. This
is a significantly stronger effect than the enhancement due to mixing
alone \cite{fkmos}. The enhancement found in this case can be
partially traced to the assumption of equal scalar masses at the weak
scale.  In particular, sfermion mixing is sensitive to this assumption
as can be seen from the magnitude of $\sin^2 (2\theta_f)$; for $m_f
\mfb \ll M_L^2 - M_R^2 + 2 Q_f \cos 2\beta \sin^2\!\theta_W M_Z^2$,
\begin{equation}
  \label{sin2thf}
  \sin^2(2\theta_f)\approx {m_f^2\mfb^2\over
                 (M_L^2 - M_R^2 + 2\,Q_f \cos 2\beta\sin^2\!\theta_W M_Z^2)^2}
\end{equation}
The assumption of equal masses translates into taking $M_L = M_R$ and hence the
enhancement in  $\sin^2 (2\theta_f)$.  Constraints from bounds on the neutron
electric dipole moment, however, restrict the range of the relevant phases to
$\thm<\pi/10$ and
$\gamma_{\rm down}<\pi/6$, so that the
$\mb(\rm max)$ is reduced to about $350\gev$. It should be noted however, that
the constraints from the neutron electric dipole moment were based on the naive
quark model and may in fact be overly restrictive when the strangeness content
of the nucleon is taken into account \cite{ef}.

The above mass ans\"{a}tz of equal masses at the weak scale is
particularly simple.  However, this pattern of low-energy sfermion
masses contains potentially dangerous charge and colour-breaking
minima in the scalar potential \cite{fkorsi,rr}.  To look at the
scalar potential above the electroweak scale, one must RGE evolve the
mass$^2$ parameters up to the scale of interest.  In the MSSM, the
sfermion mass$^2$ parameters tend to run negative at large scales due
to sfermion couplings to gauginos, and for some initial low-energy
values of the scalar and gaugino masses, CCB minima appear at high
scales.  These CCB minima may be avoided by taking the low-energy
sfermion masses to be sufficiently high.  However, these large
sfermion masses lead to a neutralino relic density which is much too
large, unless both the common $A$ parameter is tuned so that the
lighter of the two stops has a mass close to the neutralino mass and
the neutralino is heavier than the top.  In contrast, the CMSSM is
free of these minima by construction (although other CCB minima may be
present \cite{klnq}).  In this case the sfermion mass$^2$ parameters
are taken non-negative and equal at the unification scale $M_X$ and
are driven more positive by the gaugino couplings as they are run down
to the electroweak scale.  In a fully consistent model, one of the
Higgs mass$^2$ parameters runs negative at low scales to provide
$SU(2)\times U(1)$ breaking.

As we have said above, in addition to its simplicity, the CMSSM also
enjoys the feature that in most of the parameter space, the lightest
neutralino $\chi_1^0$ (and in fact the LSP) is mostly
bino\cite{susygut}, and for large values of the unified gaugino mass
$M$, $\chi_1^0$ is almost pure bino.  We can thus compute the
neutralino annihilation rate including only fermion anti-fermion final
states (as in (\ref{af})), and the error made in not considering
higgsino mixing will be insignificant for the relevant gaugino masses.
In Figure~1, we show contours of constant neutralino relic density, in
the $M$-$m_0$ plane.  Throughout this paper we use $\tan\beta=2.1$ and
$m_{\rm top}=170\gev$. Three contours are shown, for $\ohsq$ of 0.25,
0.5, and 1.  The two vertical contours are bino purity contours of
0.95 and 0.99; we see that for bino masses near their upper bounds for
$\ohsq<0.25$, where $\mb\approx0.4M\approx160 \gev$,
the higgsino admixture in the LSP is extremely small.  The shaded
regions in Figure~1 are ruled out because they lead either to a
chargino with a mass less than $\sim 65\gev$ \cite{lep1.5}, a sfermion with
a mass less than $74\gev$\cite{alitti}, or a stop, chargino, or (in
the bottom right region) stau as the LSP.  The contours of $\ohsq$
were computed for the particular choices of $A_0=300\gev, \thm=0,
\tha=0.8\,\pi$; however, the contours vary little as these parameters
are changed, and the variation is already completely negligible at the
upper range of the allowed bino mass for $\ohsq=0.25$.  There is a
similar picture for $\thm=\pi \;(\mu\rightarrow -\mu)$.

In contrast to the model studied in \cite{fkos}, the upper bound on
the bino mass in the CMSSM is essentially independent of the
$\cp$-violating phases $\thm$ and $\tha$.  Recall from (\ref{af}) that
the phases are only important if there is a significant amount of L-R
sfermion mixing.
In the CMSSM, when $M_L^2$ and $M_R^2$ are run down from $M_X$, they
are split by roughly $0.4 M^2$ at the electroweak scale. From an examination
of Eq. (\ref{sin2thf}), one can see that the off-diagonal components of the
sfermion mass matrices which contain powers of the fermion mass
 are much less than
$M_L^2-M_R^2$ when $M$ is a few hundred $\gev$, therefore sfermion mixing is
negligible, and the phases cannot lift the p-wave suppression.  Additionally,
since at the upper range of bino masses we are deep in the pure bino region,
the composition of the lightest neutralino is independent of $\mu$,
and $\thm$ in particular.  One can therefore set an upper bound for
$M$ from cosmological considerations which is quite independent of
parameter choices.  (If one takes $\tan\beta$ very large so that stau
mixing is significant, then the neutron and electron electric dipole
moments given below become very large as well.  Constraints on the
$\cp$-violating phases from bounds on EDM's then ensure that
effect of the phases on the neutralino relic density are negligible.)
One finds that for $\ohsq<0.25, M$ must be less than about $400\gev$,
which corresponds to a bino mass of roughly $160\gev$.

We turn now to the calculation of the electric dipole moments of the
neutron and the electron.  As our calculation of the EDM's parallels our
previous calculation \cite{fkos}, we refer the reader there for details.
As in \cite{fkos}, we will only include the three contributions coming from
neutralino, chargino, and gluino exchange to
the quark electric dipole moment.
  The necessary
$C\!P$ violation in these contributions comes from either
$\gf$ in the sfermion mass matrices or $\theta_\mu$ in the neutralino and
chargino mass matrices.  Full expressions for the chargino, neutralino and
gluino exchange contributions are found in \cite{ko}.

The contributions to the quark electric dipole moments from the
individual gaugino exchange diagrams fall as $M$ is increased, because
the squark masses$^2$ receive large contributions proportional to
$M^2$ during their RGE evolution from $M_X$ to $M_Z$.  Roughly,
\begin{equation}
  \label{msf}
  m_{\widetilde q}^2 \approx m_0^2 + 6 M^2 + O(M_Z^2).
\end{equation}
Thus even for large values of the $C\!P$ violating
phases, one can always turn off the quark electric dipole moment
contributions to the neutron EDM by making $M$ sufficiently large\cite{ko};
however one must still satisfy the cosmological bounds discussed
above.  Experimental bounds are $|d_n| < 1.1
\times10^{-25}e\:{\rm cm}$ \cite{nexp} for the neutron electric dipole
moment and $|d_e| < 1.9\times10^{-26}e\:{\rm cm}$ \cite{eexp}
for the electron EDM.  Note also that the squark masses $m_{\widetilde q}^2$
are only
weakly dependent on $m_0$ in the cosmologically allowed regions of
Figure~1, and so the quark EDM's will also be independent of $m_0$.

We have computed the neutron EDM in the CMSSM as a function of $\thm,
\tha,$ and $M$ for fixed $A_0, m_0, $ and $\tan\beta$, using the
na\"{\i}ve quark model.  In practice we find that the dominant
contribution to the quark electric dipole moments in the CMSSM come
from the chargino exchange diagrams, unless $\thm$ is extremely small
(and $\tha\gg\thm$).  We can then find, as a function of $\thm$ and
$\tha$, the minimum value of $M$ required to bring the quark electric
dipole moment contributions to the neutron EDM below the experimental
limits.  In Figure~2, we plot contours of $M_{\rm min}$ for
$A_0=300\gev$ and $m_0=100\gev$.  The light central region corresponds
to $M_{\rm min}<200\gev$, and successive contours represent steps of
$100\gev$.  The darkest regions on the left and right sides of the
figure lead to a stau as the LSP and correspond to the lower right
shaded region in Figure~1; their positions do depend somewhat on
$m_0$, as can be seen from Figure~1.  The EDM's were computed on
a grid with spacings of $\pi/10$ in $\tha$ and $\pi/50$ in $\thm$, so
contour features smaller than these dimensions are not significant.
Recalling from Figure~1 that
$\ohsq<0.25$ requires $M<400\gev$, we find the constraint that
$|\thm|\la\pi/15$ (the region with $\thm < 0$ comes from values of
$\tha$ near $3\pi/2$).  There is a similar allowed region near
$\thm=\pi$, corresponding to a sign change in $\mu$.  The dependence
on $\tha$ in Figure~2 is weak because it affects only the gluino
exchange contribution, which is sub dominant, and even then $\tha$ is
only partially responsible for the relevant phases $\gamma_{\rm up}$
and $\gamma_{\rm down}$ (see Eq. (\ref{mfbar})).  The figure is
shifted slightly to one side of $\thm=0$ because in one direction
there is a cancellation between the chargino and gluino exchange
pieces, and in the other direction the two contributions come in with
like signs.  These bounds are not dependent on $m_0$ and are only very
weakly dependent on $A_0$ for $A_0< 1\tev$.

In ref \cite{fkos} it was found that parameter choices which produced
a sufficiently small neutron EDM also produced an electron EDM well
below experimental bounds, so that it was never necessary to
separately consider bounds from the electron EDM.  In the case of the
CMSSM, this is not always the case.  The reason is that in the CMSSM,
the squarks receive a large contribution to their masses from the RGE
running and are consequently much heavier than the sleptons, in
contrast to the case studied in ref \cite{fkos} where all the
sfermions shared a common scalar mass$^2$ parameter at $M_Z$.  The
analysis for the electron EDM is slightly complicated by the fact that
the slepton masses, and consequently the electron EDM, do depend
somewhat on $m_0$ in the cosmologically allowed region, even at the
upper limit of the neutralino mass, where $M$ takes its largest
allowed value.  However, for values of $m_0$ below $100\gev$, this
dependence is not strong, and the minimum values of $M$ we find change
by less than $10\%$ for values of $m_0<100\gev$.  In Figure~3 we show
contours of constant $M_{\rm min}$ for $A_0=300\gev$ and
$m_0=100\gev$, with the requirement that $|d_e| <
1.9\times10^{-26}e\:{\rm cm}$ \cite{eexp}.  Again, the darkest regions on the
left and right sides of the plot lead to a stau as the LSP and
correspond to the lower right shaded region in Figure~1.  This time
the slight dependence on $\tha$ comes from the dependence of the
neutralino exchange contribution (which in this case is significant)
on the phase $\tha$.  We observe that we get comparable bounds to
those from the neutron EDM in Figure~2, $|\thm|\la\pi/10$.  It is
important to note however, that the bounds coming from the electron
electric dipole moment are clearly not sensitive to the spin structure
of the nucleon \cite{ef}.

In summary, we have combined cosmological bounds on gaugino masses with
experimental bounds on the neutron and electron electric dipole moments
to constrain the new $\cp$-violating phases in the Constrained Minimal
Supersymmetric Standard Model.  We find that in contrast to models
studied previously, the phases do not affect the cosmological limits on
the mass of an LSP bino-type neutralino.   While
there is no bound on the phase $\tha$ of the unified trilinear
scalar mass parameter $A$, the phase of the
supersymmetric Higgs mixing mass is constrained by $|\thm|\la\pi/10$.

\vskip 1in
\vbox{
\noindent{ {\bf Acknowledgments} } \\
\noindent  We would like to thank Mark Srednicki for
continual discussions.  This work was supported in
part by DOE grant DE--FG02--94ER--40823.}

%\newpage

\newpage
\vskip 2in
\noindent{\bf{Figure Captions}}

\vskip.3truein

\begin{itemize}
 \item[]
\begin{enumerate}
\item[]
\begin{enumerate}

\item[Fig.~1)] Contours of constant $\ohsq=0.25, 0.5,$ and $1.0$,
as a function of $m_0$ and $M$, for $A_0=300\gev, \tha=0.8\,\pi,
\thm=0,$ and $\tan\beta=2.1$.  The vertical lines represent contours
of constant bino purity, $p=0.95$ and $p=0.99$.  The shaded regions
yield an LSP which is either a chargino or a sfermion.

\item[Fig.~2)] Contours of constant $M_{\rm min}$, the minimum gaugino
mass parameter needed to bring the neutron electric dipole moment
below experimental bounds, for  $A_0=300\gev, m_0=100\gev$.  The light
central region corresponds to $M_{\rm min}<200\gev$, and successive
contours represent steps of $100\gev$.  Note that $M<400\gev$ implies
$\thm\la\pi/15$.

\item[Fig.~3)] Same as Fig.~2 for the electron electric dipole moment.
  In this case $M<400\gev$ implies $\thm\la\pi/10$.

\end{enumerate}
\end{enumerate}
\end{itemize}
%\newpage

\newpage

\begin{figure}[p]
\epsffile{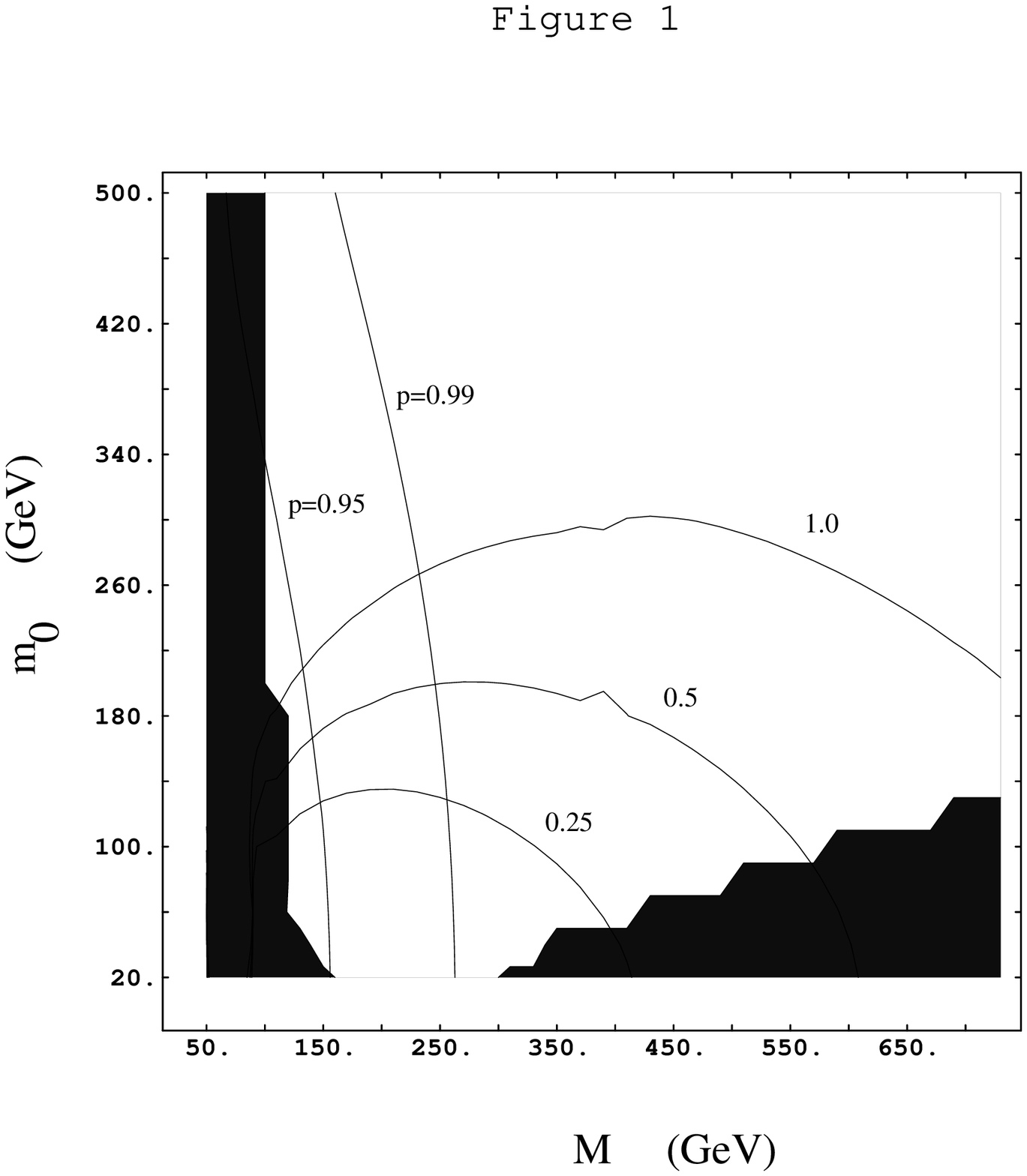}
\end{figure}
\newpage
\begin{figure}[p]
\epsffile{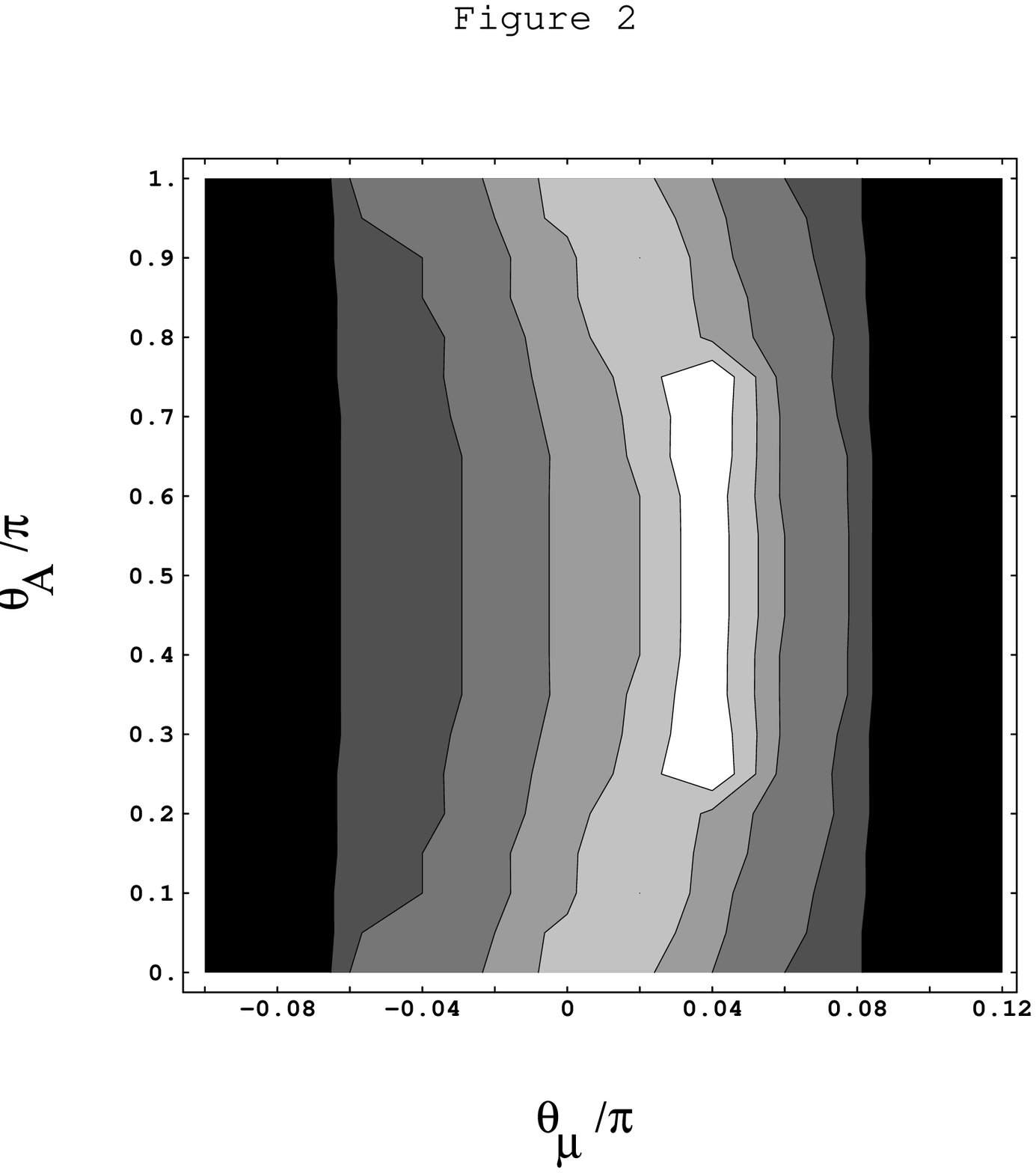}
\end{figure}
\newpage
\begin{figure}[p]
\epsffile{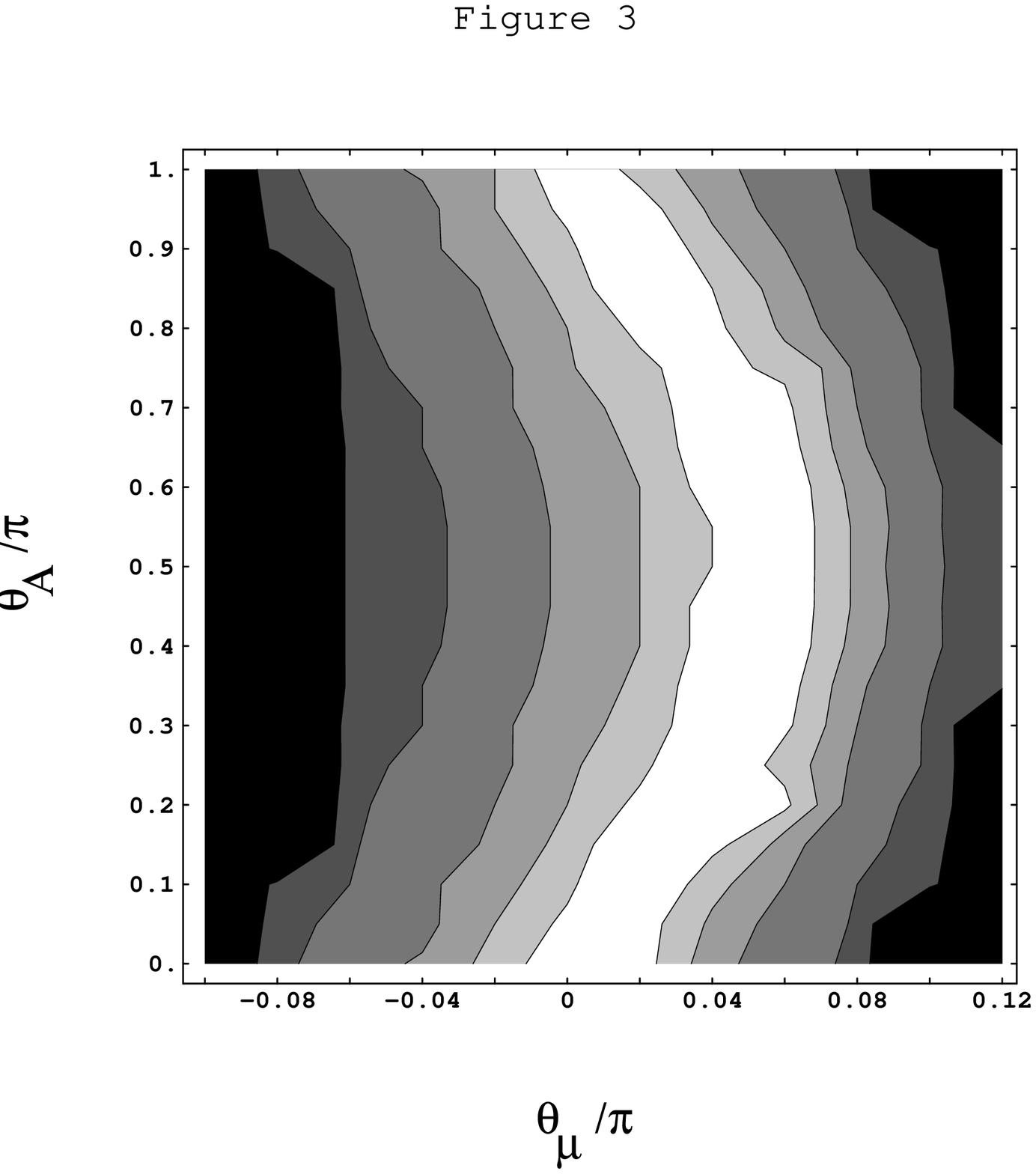}
\end{figure}

\end{document}